\documentclass[letterpaper]{article} 
\usepackage{aaai2027}  
\usepackage[hyphens]{url}  
\usepackage{graphicx} 
\usepackage{natbib}  
\usepackage{caption} 
\usepackage{algorithm}
\usepackage{algorithmic}
\usepackage{amssymb}
\usepackage{amsmath}
\usepackage{multirow}
\usepackage{arydshln}
\usepackage{newfloat}
\usepackage{listings}
\DeclareCaptionStyle{ruled}{labelfont=normalfont,labelsep=colon,strut=off} 
\floatstyle{ruled}
\newfloat{listing}{tb}{lst}{}
\floatname{listing}{Listing}

\usepackage{booktabs}

\title{HyperFL: Query-Adaptive Representation Learning for Software Fault Localization}
\author{
    Shuai Shao\textsuperscript{\rm 1},
    Yiming Zeng\textsuperscript{\rm 1},
    Yu Zhao\textsuperscript{\rm 2},
    Tingting Yu\textsuperscript{\rm 1}
}

\affiliations{
    \textsuperscript{\rm 1}University of Connecticut\\
    \textsuperscript{\rm 2}University of Cincinnati\\
    shuai.shao@uconn.edu, yiming.zeng@uconn.edu, zhao3y3@ucmail.uc.edu, tingting.yu@uconn.edu
}

\begin{document}

\maketitle

\begin{abstract}
Software fault localization identifies the code locations responsible for reported issues and is a fundamental step toward automated debugging and program repair. Recent retrieval-based approaches formulate fault localization as a dense retrieval task by learning a shared embedding space between issue reports and source code. However, these methods encode all issue reports using a fixed query representation, despite the substantial diversity of real-world issue reports in length, structure, and debugging information. To address this limitation, we propose HyperFL, a query-adaptive representation learning framework for software fault localization. HyperFL employs a lightweight hypernetwork to generate query-specific LoRA parameters for the query encoder, enabling dynamic query adaptation while keeping the code encoder fixed and reusable. Experiments on a real-world issue localization benchmark demonstrate that HyperFL consistently improves retrieval performance across multiple embedding backbones, achieving up to 13.3\% relative improvement in function-level MRR@10 and 16.7\% relative improvement in Hit@1 over the state-of-the-art method SweRank. Further analysis shows that HyperFL learns distinct adaptation patterns for different issue characteristics, highlighting the effectiveness of query-adaptive representations for software issue localization.
\end{abstract}

\section{Introduction}

Software fault localization aims to identify the code locations responsible for reported software issues and is a fundamental step toward automated debugging, program repair, and intelligent software engineering. Accurate localization allows downstream systems to focus their reasoning and modification efforts on a small subset of relevant code, thereby reducing the search space and improving repair effectiveness. As modern software systems continue to grow in scale and complexity, automated fault localization has become increasingly important, particularly for AI-based software engineering agents that rely on effective repository navigation and code understanding.

Fault localization has traditionally been studied through spectrum-based and information retrieval (IR) techniques. Spectrum-based methods identify suspicious code entities using execution information collected from test cases, whereas IR-based approaches formulate issue localization as a retrieval problem by matching natural-language issue reports with source code.
More recently, advances in large language models (LLMs) have enabled agent-based localization systems that iteratively explore repositories, inspect code, and refine their predictions through reasoning and tool interaction. Although these methods have demonstrated promising localization capabilities~\cite{autofl,agentfl,flexfl}, they typically require multiple rounds of LLM inference, resulting in substantial computational cost and latency. Their multi-step pipelines may also be sensitive to intermediate errors, as an incorrect search direction or incomplete code observation can propagate to the final prediction.

Retrieval-based approaches therefore remain attractive because of their simplicity, efficiency, and scalability. Benefiting from recent advances in code representation learning, task-specific retrievers have achieved strong localization performance while requiring only a single retrieval pass.

However, real-world software issue reports exhibit substantial diversity in both content and structure. Some reports are concise feature requests that primarily describe expected functionality, whereas others contain rich debugging evidence such as stack traces, error messages, execution environments, or code snippets. These heterogeneous signals differ considerably in both semantics and importance, making a single shared query representation unlikely to capture the most informative aspects of every issue. Nevertheless, existing retrieval-based methods typically encode all issue reports using the same query encoder and shared parameters.

Ideally, a retrieval model should adapt its query representation to each issue report while preserving a shared code representation for efficient retrieval. Motivated by this insight, we propose \textbf{HyperFL}, a lightweight query-adaptive retrieval framework for software fault localization. Rather than maintaining multiple specialized encoders or predefined issue categories, HyperFL employs a hypernetwork to generate lightweight adaptation parameters conditioned on each issue report. This design enables the query encoder to dynamically emphasize different semantic patterns while keeping the code encoder frozen and reusable. As a result, HyperFL improves localization accuracy with minimal additional computation.

We evaluate HyperFL on a real-world issue localization benchmark constructed from GitHub issues, which exhibits substantially greater diversity than existing benchmarks. Experimental results show that HyperFL consistently improves retrieval performance across multiple embedding backbones, achieving relative improvements of 13.3\% in function-level MRR and 16.7\% in Hit@1 over the state-of-the-art SweRankEmbed~\cite{swerank}. Further analyses show that HyperFL learns query-specific adaptations that particularly benefit issue reports containing richer debugging information and specialized semantics.

Our contributions are summarized as follows:

\begin{itemize}
    \item We identify the limitations of fixed query representations for retrieval-based software fault localization and motivate query-adaptive representation learning.

    \item We propose HyperFL, a lightweight framework that employs a hypernetwork to generate query-specific LoRA adaptations while preserving a shared and reusable code representation.

    \item We construct a real-world issue localization benchmark with substantially more diverse issue reports than existing benchmarks, and demonstrate consistent improvements across multiple pretrained retrieval backbones.

    \item We analyze the learned query adaptations, showing that HyperFL captures distinct adaptation patterns for different issue characteristics and consistently improves retrieval performance.
\end{itemize}
\section{Related Work}

\subsection{Software Fault Localization}

Software fault localization (FL) aims to identify the code locations responsible for observed software failures. Traditional FL techniques primarily rely on program execution information. Spectrum-based fault localization ranks program elements according to their statistical association with passing and failing test executions~\cite{sp1,sp2,sp3}, while mutation-based methods estimate suspiciousness by analyzing how program mutations affect test outcomes~\cite{m1,m2}. Static and dynamic program analyses, including call-graph traversal, dependency analysis, and program slicing, have also been used to constrain the search space of potential fault locations~\cite{yu2008empirical, shao2023information,blizzard,shao2024enhancing,BTrace}. Although effective in controlled settings, these methods generally require reliable test cases, execution traces, or precise program models, which may not be available for real-world software issues.

Information-retrieval-based FL instead treats a bug report as a natural-language query and ranks source-code entities according to textual or semantic similarity. These methods can directly exploit issue descriptions without requiring executable tests, but classical lexical approaches depend heavily on vocabulary overlap between bug reports and source code and often perform poorly at fine-grained localization levels. More recent learning-based methods employ pretrained code models to learn aligned representations of bug reports and source code~\cite{rank1,rank2, rank3,swerank}. In particular, SweRank~\cite{swerank} formulates issue localization as a retrieve-and-rerank problem and trains a task-specific code retriever using issue--function pairs collected from GitHub. However, these approaches still use a single fixed query encoder for all issue reports, despite substantial variation in their structure and information content.

The emergence of large language models has inspired a new line of research in fault localization. Systems such as AutoFL~\cite{autofl} and AgentFL~\cite{agentfl} perform iterative reasoning over bug reports and program context, while FlexFL~\cite{flexfl} further combines bug report information with test executions. Although these systems achieve promising localization performance, they typically require multiple rounds of reasoning and repository exploration, resulting in substantial inference cost and latency. Moreover, their sequential decision process can be brittle, as incorrect intermediate decisions or incomplete code observations may propagate through subsequent reasoning steps and ultimately degrade localization accuracy.

\subsection{Code Retrieval and Adaptive Representation Learning}

Transformer-based code retrieval models learn a shared embedding space between natural-language queries and source code and have achieved strong performance on tasks such as text-to-code search and code-to-code retrieval~\cite{codebert,graphcodebert,unixcoder}. These models are commonly trained using language-modeling, text-code matching, and contrastive objectives over large-scale code corpora~\cite{codebert,unixcoder,rank1}. Recent retrieval-oriented models further improve representation quality through large-scale synthetic supervision, hard-negative mining, and retrieval-specific training objectives~\cite{swerank}.

Parameter-efficient adaptation methods provide a lightweight mechanism for specializing pretrained models. LoRA introduces low-rank parameter updates while keeping the original model weights fixed~\cite{lora}, whereas hypernetworks generate the parameters of another network conditioned on an input or task representation~\cite{ha2017hypernetworks}. Prior work has successfully applied hypernetworks across diverse application domains, demonstrating their ability to generate adaptive model parameters for different inputs~\cite{h1,h2,h3}. In contrast, HyperFL performs \emph{query-level} adaptation: for each issue report, a lightweight hypernetwork generates query-specific LoRA parameters for the query encoder. This allows the model to construct representations that adapt to the characteristics of individual issues while preserving a fixed and reusable code representation space.
\section{Method}
\label{sec:method}

\begin{figure*}[t]
    \centering
    \includegraphics[width=0.9\linewidth]{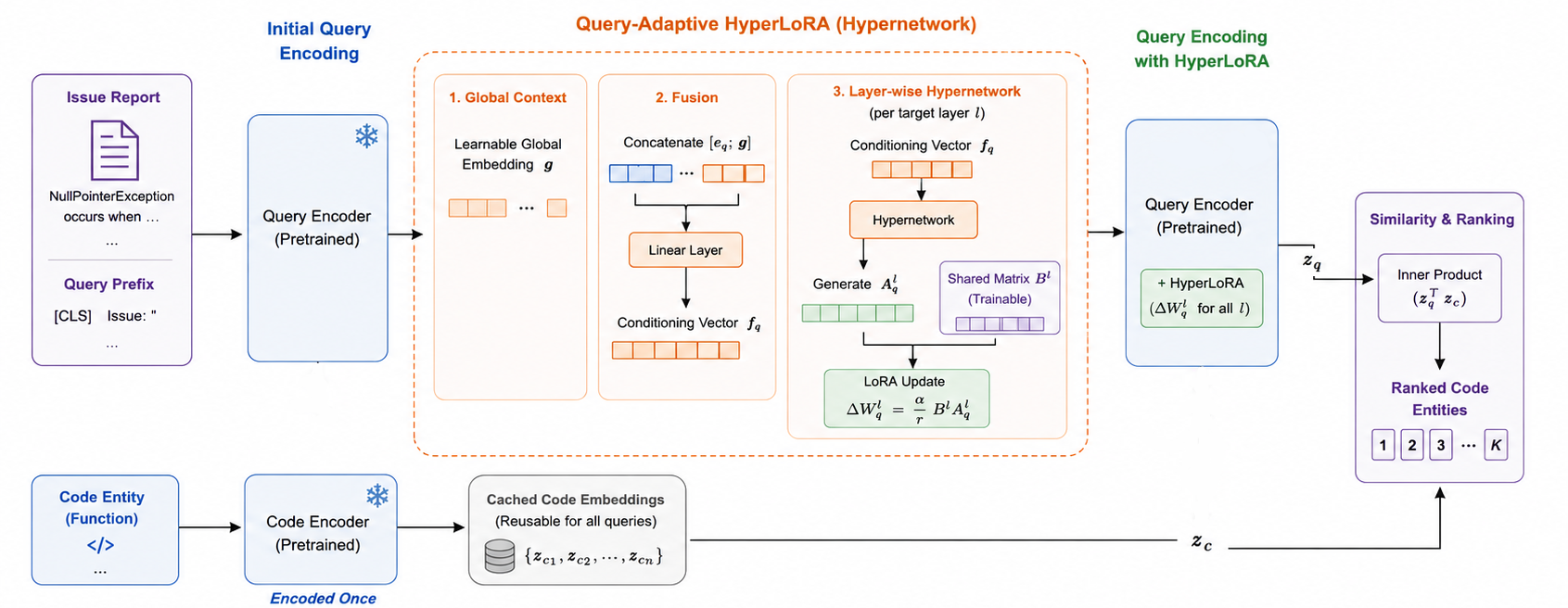}
    \caption{Overview of the HyperFL framework.}
    \label{fig:hyperfl-overview}
\end{figure*}

\subsection{Problem Formulation}
\label{sec:problem}

Given an issue report $q$ and a repository represented by candidate code entities
$\mathcal{C}=\{c_1,\ldots,c_n\}$,
software fault localization aims to rank code entities according to their relevance to the reported issue. We formulate this task as dense retrieval between issue reports and code entities.

A conventional dense retriever adopts a bi-encoder architecture consisting of a query encoder $f_q$ and a code encoder $f_c$:
\begin{equation}
\mathbf{z}_q=f_q(q;\theta_q), \qquad
\mathbf{z}_c=f_c(c;\theta_c),
\label{eq:static-encoders}
\end{equation}
where $\theta_q$ and $\theta_c$ denote the encoder parameters. Candidate code entities are ranked according to the cosine similarity
\begin{equation}
s(q,c)=
\frac{\mathbf{z}_q^\top\mathbf{z}_c}
{\|\mathbf{z}_q\|_2\|\mathbf{z}_c\|_2}.
\label{eq:retrieval-score}
\end{equation}

Existing retrieval models use a single shared query encoder for all issue reports, despite their diverse characteristics. HyperFL instead adopts a query-adaptive representation:
\begin{equation}
\mathbf{z}_q=
f_q\!\left(q;\theta_q+\Delta\theta(q)\right),
\label{eq:adaptive-query}
\end{equation}
where $\Delta\theta(q)$ is a query-specific low-rank adaptation generated for the current issue report. The pretrained query encoder remains frozen, while the code representation is computed as
\begin{equation}
\mathbf{z}_c=f_c(c;\theta_c),
\label{eq:fixed-code}
\end{equation}
using a fixed code encoder. This asymmetric design enables query-specific adaptation while preserving a shared code representation space that can be precomputed and reused for efficient retrieval.


\subsection{Framework Overview}
\label{sec:overview}

Figure~\ref{fig:hyperfl-overview} presents the overall architecture of HyperFL. Given an issue report, HyperFL performs two query-encoding passes. In the first pass, the frozen query encoder produces an initial query representation, which serves as the conditioning signal for the HyperLoRA module. The module first augments the query representation with a learnable global context, projects the fused representation into a conditioning vector, and then employs a layer-wise hypernetwork to generate query-specific LoRA parameters for the selected layers of the query encoder.

In the second pass, the issue report is re-encoded using the generated HyperLoRA parameters, producing the final query embedding for retrieval. In contrast, candidate code entities are encoded only once by the frozen code encoder. Their embeddings are cached and reused across all queries, enabling efficient retrieval without repeatedly encoding the repository.

Finally, the adapted query embedding is compared with the cached code embeddings to rank candidate code entities. This asymmetric design combines query-specific adaptation with a shared and reusable code representation space.

\subsection{Initial Query Conditioning}
\label{sec:conditioning}

HyperFL first obtains an initial representation of each issue report using the frozen query encoder. Following the input format of the underlying embedding model, we prepend a retrieval instruction to the issue report and encode the resulting query without activating HyperLoRA:
\[
\widetilde{q}=[\texttt{query-prefix};q],
\qquad
\mathbf{e}_q=
\operatorname{Norm}\!\left(
f_q(\widetilde{q};\theta_q)
\right),
\]
where $\mathbf{e}_q\in\mathbb{R}^{d}$ is the normalized representation of the first token in the final hidden layer.

To provide shared task-level information, HyperFL maintains $K$ learnable global embeddings
$G=[\mathbf{g}_1,\ldots,\mathbf{g}_K]\in\mathbb{R}^{K\times d}$,
which are summarized by a GRU:
\begin{equation}
\mathbf{g}
=
\operatorname{GRU}_{\mathrm{global}}(G)_K,
\label{eq:global-context}
\end{equation}
where $\mathbf{g}\in\mathbb{R}^{d}$ denotes the global context vector.

The initial query representation and global context are then fused through a linear projection:
\begin{equation}
\mathbf{f}_q
=
W_f[\mathbf{e}_q;\mathbf{g}]
+
\mathbf{b}_f,
\label{eq:context-fusion}
\end{equation}
where $\mathbf{f}_q\in\mathbb{R}^{d}$ serves as the conditioning vector for generating query-specific HyperLoRA parameters across all target layers.


\subsection{Query-Adaptive HyperLoRA}
\label{sec:hyperlora}

Given the conditioning vector $\mathbf{f}_q$, HyperFL generates query-specific LoRA parameters for each target transformation of the query encoder. Rather than sharing a single parameter generator across all transformer layers, HyperFL adopts a layer-wise hypernetwork to capture the distinct adaptation requirements of different layers. The generation process consists of three stages: (1) layer-specific conditioning, (2) query-specific LoRA generation, and (3) dynamic weight updates.

\subsubsection{Layer-Specific Conditioning}

Different transformer layers capture different levels of lexical, syntactic, and semantic information. HyperFL therefore maintains an independent GRU for each target transformation. For each target layer
$l\in\mathcal{L}$,
the conditioning vector is transformed as
\begin{equation}
\mathbf{h}_q^l
=
\operatorname{GRU}_l(\mathbf{f}_q),
\label{eq:layer-gru}
\end{equation}
where
$\mathbf{h}_q^l\in\mathbb{R}^{d_h}$ denotes the layer-specific conditioning vector.

\subsubsection{Query-Specific LoRA Generation}

Each target transformation additionally maintains a learnable local embedding
\begin{equation}
E^l
\in
\mathbb{R}^{d_{\mathrm{in}}^l\times r},
\end{equation}
which is shared across all issue reports. The layer-specific conditioning vector is broadcast along the input dimension and concatenated with the local embedding:
\begin{equation}
Z_q^l
=
[E^l;\overline{H}_q^l].
\label{eq:local-dynamic-fusion}
\end{equation}

A lightweight projection network then generates the query-specific LoRA matrix
\begin{equation}
A_q^l
=
P_l(Z_q^l),
\qquad
A_q^l
\in
\mathbb{R}^{r\times d_{\mathrm{in}}^l},
\label{eq:a-generation}
\end{equation}
where $P_l(\cdot)$ consists of layer normalization, GeLU activation, dropout, and a linear projection.

HyperFL generates only the LoRA matrix $A_q^l$. The complementary matrix
\begin{equation}
B^l
\in
\mathbb{R}^{d_{\mathrm{out}}^l\times r}
\label{eq:static-b}
\end{equation}
is trainable but shared across all issue reports. This decomposition separates reusable transformation knowledge from query-specific adaptation.

\subsubsection{Query-Specific Weight Update}

The generated low-rank matrices define the query-specific LoRA update:
\begin{equation}
\Delta W_q^l
=
\frac{\alpha}{r}
B^lA_q^l,
\label{eq:dynamic-update}
\end{equation}
where $\alpha$ is the LoRA scaling factor.

For an input activation
$X_q^l\in\mathbb{R}^{T\times d_{\mathrm{in}}^l}$,
the adapted transformation becomes
\begin{align}
Y_q^l
&=
X_q^l(W^l+\Delta W_q^l)^\top \nonumber\\
&=
X_q^l(W^l)^\top
+
\frac{\alpha}{r}
X_q^l(A_q^l)^\top(B^l)^\top.
\label{eq:adapted-forward}
\end{align}

The pretrained weight matrix $W^l$ remains frozen throughout training, while the generated matrix $A_q^l$ varies across issue reports. Consequently, different queries produce different LoRA updates, enabling HyperFL to dynamically adapt the query encoder while preserving the pretrained embedding model.

\subsection{Asymmetric Query and Code Encoding}
\label{sec:asymmetric}

After generating the query-specific HyperLoRA parameters, HyperFL performs asymmetric encoding for issue reports and code entities. The query encoder applies the generated LoRA updates, whereas the code encoder remains frozen.

The adapted query representation is computed as
\begin{equation}
\mathbf{z}_q
=
\operatorname{Norm}
\left(
f_q
\left(
\widetilde{q};
\theta_q,
\{A_q^l,B^l\}_{l\in\mathcal{L}}
\right)
\right),
\label{eq:final-query}
\end{equation}
where the generated HyperLoRA parameters are activated only during query encoding. In contrast, the code representation is obtained using the frozen code encoder:
\begin{equation}
\mathbf{z}_c
=
\operatorname{Norm}
\left(
f_c(c;\theta_c)
\right).
\label{eq:final-code}
\end{equation}

The final retrieval score is computed as
\begin{equation}
s(q,c)=\mathbf{z}_q^\top\mathbf{z}_c,
\label{eq:final-similarity}
\end{equation}
where both representations are L2-normalized. Since only the query encoder is dynamically adapted, repository code embeddings can be precomputed and reused across all issue reports, preserving the efficiency of dense retrieval.


\subsection{Training}
\label{sec:training}

Each training instance consists of an issue report, its corresponding positive code entity, and a set of negative code entities. Following prior dense retrieval methods, negative candidates are first ranked by retrieval difficulty. To reduce the influence of mislabeled or near-positive samples, we discard the hardest candidates and randomly sample from the remaining pool, resulting in informative medium-hard negatives.

HyperFL is trained using the InfoNCE objective:
\begin{equation}
\mathcal{L}
=
-
\log
\frac{
\exp(s(q,c^+)/\tau)
}{
\exp(s(q,c^+)/\tau)
+
\sum_{c^-\in\mathcal{N}}
\exp(s(q,c^-)/\tau)
},
\label{eq:infonce}
\end{equation}
where $s(\cdot,\cdot)$ denotes the retrieval score defined in Eq.~(\ref{eq:final-similarity}) and $\tau$ is the temperature parameter.

During training, both pretrained query and code encoders remain frozen. Only the HyperFL components, including the global context, layer-wise hypernetwork, local embeddings, and shared LoRA parameters, are optimized.


\section{Experiments}
\begin{table*}[!t]
\centering
\small
\caption{Overall issue localization performance at the file and function levels.}
\label{tab:overall}
\begin{tabular}{lccccc|ccccc}
\hline
\multirow{2}{*}{\textbf{Models}} & \multicolumn{5}{c|}{\textbf{File-level}} & \multicolumn{5}{c}{\textbf{Function-level}} \\ \cline{2-11}
& \textbf{Hit@1} & \textbf{Hit@5} & \textbf{Hit@10} & \textbf{MRR} & \textbf{Acc@10} & \textbf{Hit@1} & \textbf{Hit@5} & \textbf{Hit@10} & \textbf{MRR} & \textbf{Acc@10} \\ \hline
BM25                 & 0.41 & 0.74 & 0.85 & 0.55 & 0.72 & 0.24 & 0.43 & 0.52 & 0.32 & 0.31 \\
CodeRankEmbed        & 0.40 & 0.73 & 0.83 & 0.55 & 0.72 & 0.21 & 0.40 & 0.47 & 0.29 & 0.27 \\
Jina-Code-v2         & 0.50 & 0.81 & 0.89 & 0.64 & 0.78 & 0.25 & 0.47 & 0.55 & 0.35 & 0.33 \\
Qwen3-Embedding      & 0.41 & 0.77 & 0.86 & 0.56 & 0.74 & 0.24 & 0.43 & 0.51 & 0.32 & 0.32 \\
SweRankEmbed-Small   & 0.62 & 0.87 & 0.93 & 0.73 & 0.82 & 0.36 & 0.57 & 0.69 & 0.45 & 0.44 \\ \hdashline
HyperFL (CodeRankEmbed) & 0.67 & 0.88 & 0.92 & 0.76 & 0.82 & 0.42 & 0.65 & 0.71 & 0.51 & 0.43 \\
HyperFL (Jina-Code-v2)  & \textbf{0.73} & 0.91 & \textbf{0.95} & \textbf{0.81} & \textbf{0.86} & \textbf{0.47} & \textbf{0.70} & \textbf{0.77} & \textbf{0.57} & \textbf{0.49} \\
HyperFL (Qwen3-Embedding) & 0.70 & \textbf{0.92} & 0.94 & 0.79 & 0.85 & 0.46 & \textbf{0.70} & 0.75 & 0.56 & 0.48 \\ \hline
\end{tabular}
\end{table*}
\subsection{Experimental Setup}

\textbf{Dataset.}
Following the data collection protocol of SWE-bench~\cite{jimenez2024swe}, we construct a function-level issue localization benchmark from real-world GitHub repositories. We randomly select five repositories and collect 369 issue reports together with their corresponding fixing commits. For each issue, we identify the modified functions in the fixing commit as the ground-truth localization targets, while treating all remaining functions in the repository as retrieval candidates.

\textbf{Baselines.}
We compare HyperFL against lexical, pretrained dense retrieval, and task-specific software issue localization methods.
BM25~\cite{bm25} serves as the lexical baseline.
For pretrained dense retrieval, we include Jina-Code-v2
~\cite{jina}, CodeRankEmbed~\cite{rank3},
and Qwen3-Embedding-0.6B~\cite{qwen}.
The first two are compact code-oriented embedding models, whereas
Qwen3-Embedding-0.6B provides a stronger and more recent general-purpose
embedding baseline with substantially more parameters.
We use SweRankEmbed-Small~\cite{swerank} as the primary task-specific
software issue localization baseline.

\textbf{Evaluation Metrics.}
Following prior work on software issue localization, we evaluate function-level retrieval performance using Mean Reciprocal Rank (MRR), Hit Rate (Hit@$k$), and Accuracy (ACC). MRR@10 measures the reciprocal rank of the highest-ranked ground-truth function within the top ten predictions, assigning zero when no relevant function appears in the top ten. Hit@$k$ indicates whether at least one ground-truth function is retrieved among the top $k$ candidates. ACC measures the proportion of issue reports for which the top-ranked prediction is a ground-truth function. We report MRR@10, Hit@1, Hit@5, Hit@10, and ACC. Higher values indicate better localization performance.

\textbf{Training Data Construction.}
To ensure a consistent training distribution across different retrieval backbones, we use BM25 as a backbone-independent filter to construct the training set. Unless otherwise specified, we retain issue--function pairs whose ground-truth function is ranked within the top 30 candidates returned by BM25. This filtering strategy provides a unified training set for all retrieval models while remaining independent of any particular embedding backbone. The impact of the filtering threshold and training-set size is analyzed in a later experiment.

\textbf{Implementation Details.}
HyperFL uses a LoRA rank of 8 with a scaling factor of 32. Models are optimized using the InfoNCE loss with a temperature of 0.05 and 10 negative samples per query. We use the AdamW optimizer with a learning rate of $2\times10^{-5}$, a batch size of 16, and gradient accumulation over 4 steps (effective batch size 64). The maximum input length is set to 512 tokens. Models are trained for 3 epochs, and the checkpoint with the best validation MRR@10 is selected for evaluation. All experiments are conducted on a single NVIDIA A100-SXM4-80GB GPU using a random seed of 42.

\subsection{Overall Localization Performance}

Table~\ref{tab:overall} reports the overall localization performance on our real-world benchmark at both the file and function levels. HyperFL consistently outperforms lexical retrieval, pretrained embedding models, and the state-of-the-art issue localization method SweRank across all evaluation metrics. Since SweRank is built upon the CodeRankEmbed backbone, we first compare it with HyperFL instantiated on the same encoder to isolate the effect of query-adaptive representation learning. Using the identical CodeRankEmbed backbone, HyperFL improves function-level MRR from 0.45 to 0.51 and Hit@1 from 0.36 to 0.42, corresponding to relative improvements of 14.1\% and 19.1\%, respectively. Consistent gains are also observed at the file level, demonstrating that the improvements stem from the proposed query-adaptive HyperLoRA rather than a stronger pretrained encoder.

To evaluate its generality, we instantiate HyperFL on three representative embedding models: CodeRankEmbed, Jina Code Embeddings, and Qwen3 Embedding.
HyperFL consistently improves all three backbones, increasing function-level MRR from 0.29 to 0.51 for CodeRankEmbed, 0.35 to 0.57 for Jina Code Embeddings, and 0.32 to 0.56 for Qwen3 Embedding.
The best overall performance is achieved with Jina Code Embeddings, reaching a function-level MRR of 0.57 and Hit@1 of 0.47.

Overall, these results demonstrate that query-specific adaptation consistently improves retrieval effectiveness across diverse pretrained retrievers, indicating that HyperFL complements pretrained representations rather than relying on a particular encoder architecture.

\subsection{Performance Across Diverse Issue Characteristics}
\label{sec:issue_characteristics}

Software issue reports vary substantially in both content and structure, often combining natural-language descriptions, error messages, stack traces, code snippets, and reproduction steps. Since these characteristics frequently overlap, manually defining issue categories is inherently ambiguous. Instead, we cluster the query-specific LoRA parameters generated by HyperFL and characterize each cluster post hoc using observable issue features. The clustering is performed solely on the generated adaptation parameters, without using issue features or localization outcomes.

\begin{figure}[h]
    \centering
    \includegraphics[width=0.95\linewidth]{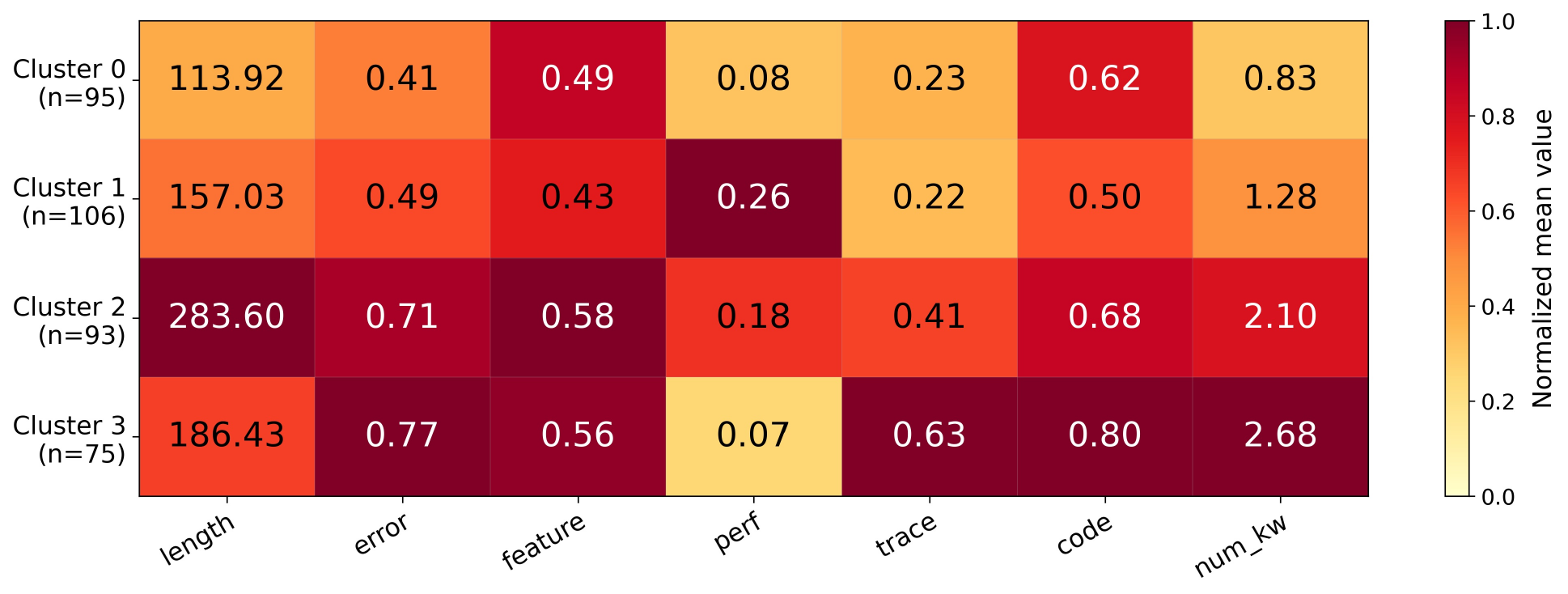}
    \caption{Issue characteristics across query clusters.}
    \label{fig:feature_heatmap}
\end{figure}

Figure~\ref{fig:feature_heatmap} summarizes the characteristics of the resulting clusters. The reported features include issue length (\textit{length}), the proportions of issue reports containing error-related keywords (\textit{error}), feature requests (\textit{feature}), performance-related keywords (\textit{perf}), stack traces (\textit{trace}), and code snippets (\textit{code}), together with the average number of error keywords (\textit{num\_kw}). Cluster~0 consists of short issue reports with relatively frequent code snippets but limited debugging information. Cluster~1 mainly contains performance-related issues, exhibiting the highest proportion of performance keywords. Cluster~2 is dominated by verbose reports containing richer textual descriptions together with substantial error and code information. Cluster~3 contains issue reports with the richest debugging context, including the highest frequency of error keywords, stack traces, and code snippets.

Table~\ref{tab:cluster_performance} reports localization performance across the four clusters. HyperFL achieves the best function-level MRR on three of the four clusters and remains competitive on Cluster~0. The largest improvement is observed on Cluster~1, where function-level MRR increases from 0.54 to 0.66, while Cluster~3 also exhibits a notable gain from 0.38 to 0.42. In contrast, improvements on Clusters~0 and~2 are relatively modest, where shorter reports or richer lexical descriptions already provide sufficient signals for existing embedding models.

These results suggest that the benefits of query-specific adaptation depend on the characteristics of the issue report. HyperFL delivers the largest improvements on issue categories that are less effectively captured by fixed embedding models, particularly performance-related reports and issues containing rich debugging context. For issue reports that are already well represented by pretrained embeddings, HyperFL maintains competitive performance without sacrificing retrieval accuracy. This demonstrates that query-adaptive representations improve robustness across diverse issue characteristics by allocating adaptation capacity where it is most needed, rather than relying on a single shared representation for all queries.

\begin{table}[t]
\centering
\caption{MRR@10 across clusters.}
\label{tab:cluster_performance}
\resizebox{\columnwidth}{!}{
\begin{tabular}{lcccccccc}
\toprule
& \multicolumn{2}{c}{\textbf{Cluster 0}}
& \multicolumn{2}{c}{\textbf{Cluster 1}}
& \multicolumn{2}{c}{\textbf{Cluster 2}}
& \multicolumn{2}{c}{\textbf{Cluster 3}} \\
\cmidrule(lr){2-3}
\cmidrule(lr){4-5}
\cmidrule(lr){6-7}
\cmidrule(lr){8-9}
\textbf{Model}
& \textbf{File} & \textbf{Func.}
& \textbf{File} & \textbf{Func.}
& \textbf{File} & \textbf{Func.}
& \textbf{File} & \textbf{Func.} \\
\midrule
CodeRankEmbed
& 0.59 & 0.32
& 0.64 & 0.42
& 0.59 & 0.23
& 0.29 & 0.13 \\

SweRankEmbed
& \textbf{0.79} & 0.52
& 0.77 & 0.54
& \textbf{0.69} & 0.33
& 0.66 & 0.38 \\

HyperFL (CodeRank)
& 0.78 & \textbf{0.55}
& \textbf{0.88} & \textbf{0.66}
& \textbf{0.69} & \textbf{0.38}
& \textbf{0.68} & \textbf{0.42} \\
\bottomrule
\end{tabular}
}
\end{table}

\subsection{Characterizing the Real-World Benchmark}

Table~\ref{tab:dataset_statistics} compares the statistics of our benchmark with SWE-bench Lite.
Compared with SWE-bench Lite, our benchmark contains longer issue reports (280.2 vs.\ 237.3 tokens on average) together with substantially larger variation in report length (431.2 vs.\ 294.1 standard deviation), indicating a broader distribution of issue descriptions.
These observations suggest that our benchmark presents a more heterogeneous retrieval scenario, which directly motivates the query-adaptive design of HyperFL.

\begin{table}[t]
\centering
\small
\caption{Issue-length statistics of the evaluation benchmarks.}
\label{tab:dataset_statistics}
\begin{tabular}{lrr}
\toprule
\textbf{Statistic} & \textbf{SWE-bench Lite} & \textbf{Real-World} \\
\midrule
Mean tokens       & 237.3 & 280.2 \\
Median tokens     & 153.5 & 180.0 \\
Std. tokens       & 294.1 & 431.2 \\
Interquartile range & 166.8 & 239.0 \\
\bottomrule
\end{tabular}
\end{table}

\subsubsection{Performance on SWE-bench Lite}

To evaluate the generality of HyperFL, we further conduct experiments on SWE-bench Lite. As shown in Table~\ref{tab:swe}, HyperFL achieves the best Hit@1 while matching SweRank in MRR, despite being trained with only approximately 30K issue--code pairs, compared with the over 100K training pairs used by SweRank.

Compared with the improvements on our benchmark, the gains on SWE-bench Lite are relatively modest. This is consistent with the findings in the previous analysis: HyperFL provides the largest benefits when pretrained embeddings struggle to represent diverse issue reports, while yielding smaller gains when the underlying embeddings are already sufficiently effective. Indeed, the CodeRankEmbed backbone achieves a considerably higher baseline MRR on SWE-bench Lite (0.37) than on our benchmark (0.29), suggesting that SWE-bench Lite contains issue reports that are more readily handled by fixed embedding representations. Consequently, the additional benefit of query-specific adaptation is naturally smaller.

Overall, these results indicate that HyperFL generalizes well to existing benchmarks while providing the greatest advantage on more diverse and challenging real-world issue localization scenarios.

\begin{table}[] \scriptsize \centering
\caption{Performance on SWE-bench Lite.}
\label{tab:swe}
\begin{tabular}{lccccc}
\hline
\textbf{Models}  & \textbf{Hit@1} & \textbf{Hit@5} & \textbf{Hit@10} & \textbf{MRR} & \textbf{Acc@10}\\ \hline
CodeRankEmbed           & 0.27 & 0.53 & 0.61 & 0.37 & 0.58 \\
SweRankEmbed            & 0.31 & \textbf{0.61} & \textbf{0.72} & \textbf{0.44} & \textbf{0.69} \\
HyperFL (CodeRank)      & \textbf{0.33} & 0.57 & 0.68 & \textbf{0.44} & 0.66 \\ \hline
\end{tabular}
\end{table}

\subsection{Impact of Training Data Construction}

\begin{figure}[t]
    \centering
    \includegraphics[width=0.9\linewidth]{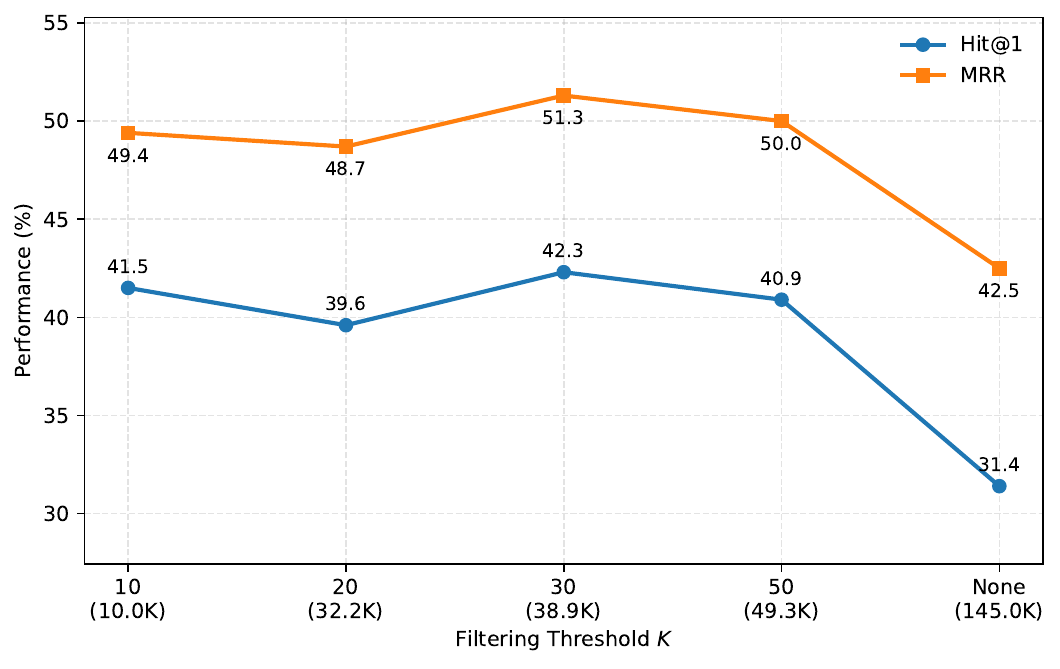}
    \caption{Impact of training data construction.}
    \label{fig:data_quality_size}
\end{figure}

HyperFL is trained using automatically constructed issue--code pairs filtered by BM25.
To investigate the trade-off between training data quality and quantity, we vary the BM25 filtering threshold $K$, retaining only issue--code pairs whose corresponding ground-truth function appears within the top-$K$ retrieved candidates.
Smaller values of $K$ produce fewer but higher-quality training pairs, while larger values introduce additional supervision at the cost of increased noise.

Figure~\ref{fig:data_quality_size} reports the localization performance under different filtering thresholds together with the resulting training set sizes.
The best performance is achieved at $K=30$, where HyperFL reaches 42.3\% Hit@1 and 51.3\% MRR using approximately 39K training pairs.
Both stricter filtering ($K=10$ or $20$) and looser filtering ($K=50$) lead to lower localization performance, indicating that neither maximizing data quality nor simply increasing training data is sufficient.

Notably, removing BM25 filtering entirely increases the training set to approximately 145K pairs but causes a substantial performance drop, reducing MRR from 51.3\% to 42.5\%.
This observation suggests that noisy issue--code pairs severely hinder learning and that effective training data construction is crucial for retrieval-based issue localization.
Unless otherwise specified, we therefore adopt $K=30$ throughout the remaining experiments.

\subsection{Analysis of Query-Specific Adaptations}

\begin{figure}[t]
    \centering
    \includegraphics[width=0.9\linewidth]{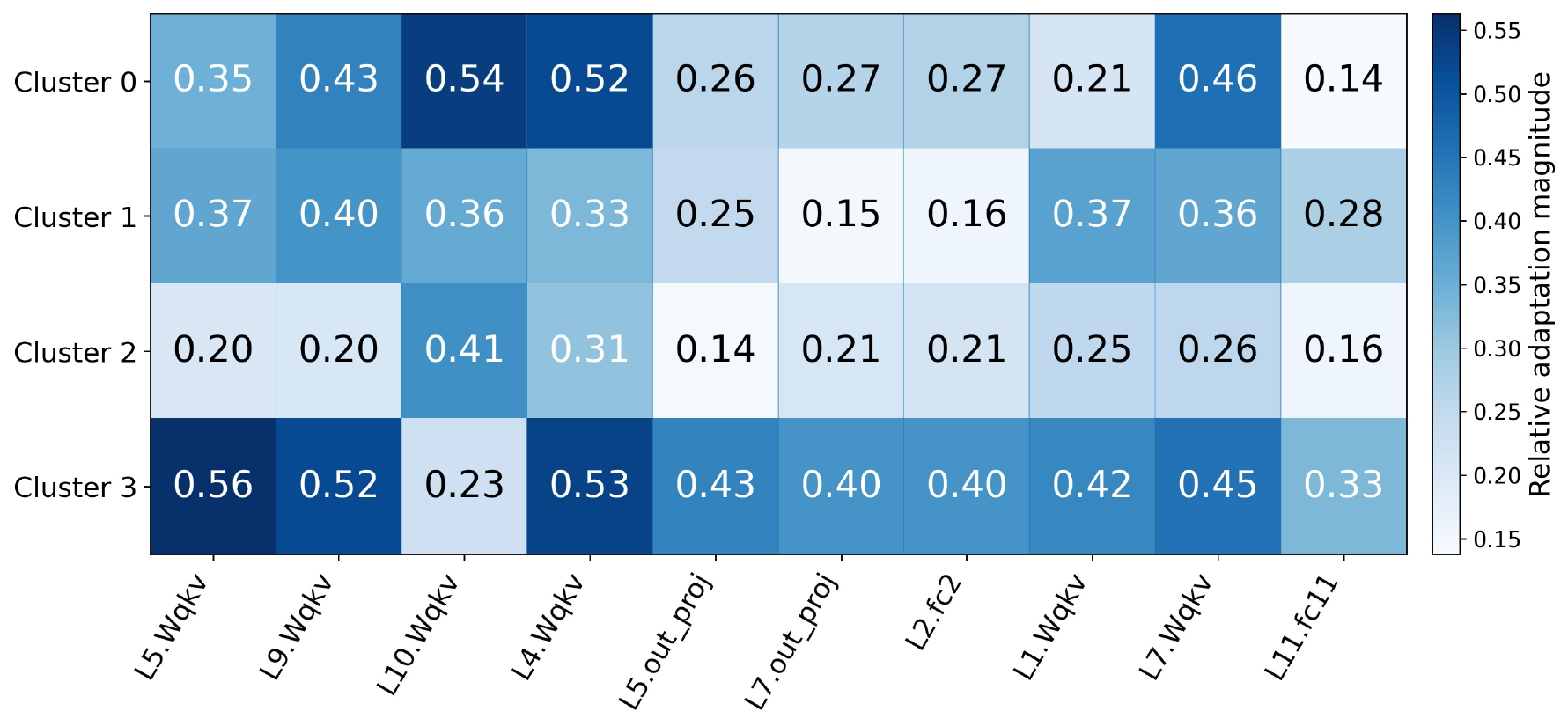}
    \caption{Layer-wise HyperLoRA adaptation across different issue clusters.}
    \label{fig:layer_adaptation}
\end{figure}

To better understand how HyperFL adapts to different issue reports, we analyze the layer-wise LoRA adaptations generated by the hypernetwork for each issue cluster. Figure~\ref{fig:layer_adaptation} reports the average adaptation magnitude on each selected transformer layer.

Distinct issue clusters exhibit substantially different adaptation patterns across transformer layers, indicating that HyperFL learns query-dependent parameter updates rather than applying a uniform modification to the encoder. Across all clusters, the largest adaptations are consistently concentrated in the attention projection matrices, particularly the WorkKV modules in the intermediate transformer layers, while the output projection layers generally receive smaller updates.

The adaptation magnitude also varies considerably across issue clusters. Cluster~3 consistently exhibits the strongest adaptations across most layers, whereas Cluster~2 receives the smallest overall updates. More importantly, the clusters emphasize different subsets of transformer layers: Cluster~0 primarily adapts intermediate layers, while Cluster~1 allocates relatively more adaptation to earlier layers. These results suggest that HyperFL not only adjusts the strength of adaptation for different issue types, but also learns where in the encoder adaptation is most beneficial, providing further evidence for query-specific representation learning.

These observations indicate that the hypernetwork learns structured layer-wise adaptation patterns conditioned on the input issue report. Combined with the performance improvements observed across issue clusters, the results suggest that HyperFL dynamically allocates model capacity to different parts of the encoder according to the semantic characteristics of each query.

\subsection{Ablation Study}

To evaluate the contribution of the hypernetwork, we remove the query-conditioned hypernetwork and use the frozen query encoder directly, resulting in a fixed query representation for all issue reports. As shown in Table~\ref{tab:ablation}, removing the hypernetwork consistently degrades localization performance across all evaluation metrics. In particular, function-level MRR decreases from 0.51 to 0.43, while Hit@1 drops from 0.42 to 0.33.

These results demonstrate that the performance gains of HyperFL primarily stem from query-specific adaptation rather than the underlying embedding model alone. By dynamically generating query-dependent LoRA parameters, the hypernetwork enables the encoder to better capture the diverse characteristics of software issue reports, leading to more discriminative retrieval representations.
\begin{table}[] \scriptsize \centering
\caption{Ablation study.}
\label{tab:ablation}
\begin{tabular}{lccccc}
\hline
\textbf{Models}  & \textbf{Hit@1} & \textbf{Hit@5} & \textbf{Hit@10} & \textbf{MRR} & \textbf{Acc@10} \\ \hline
HyperFL     &  \textbf{0.42} & \textbf{0.65} & \textbf{0.71} & \textbf{0.51} & \textbf{0.43} \\
w/o hypernetwork    & 0.33 & 0.56 & 0.65 & 0.43 & 0.39 \\ \hline
\end{tabular}
\end{table}

\section{Conclusion}

This paper presented HyperFL, a query-adaptive representation learning framework for software fault localization. Unlike existing retrieval-based approaches that encode all issue reports using a fixed query representation, HyperFL dynamically generates query-specific HyperLoRA parameters through a lightweight layer-wise hypernetwork, enabling the query encoder to adapt to diverse issue characteristics while keeping the code encoder fixed and reusable. Extensive experiments on a real-world issue localization benchmark demonstrated that HyperFL consistently improves localization performance across multiple embedding backbones and achieves state-of-the-art results. Further analyses showed that HyperFL learns distinct adaptation patterns for different issue characteristics and that high-quality training data plays a critical role in retrieval performance. These results suggest that query-adaptive representation learning is a promising direction for scalable software issue localization and other retrieval-based software engineering tasks.

\bibliography{aaai2027}

\end{document}